\documentclass[twocolumn,superscriptaddress,nofootinbib,floatfix,aps,prb,citeautoscript,longbibliography]{revtex4-2}
\usepackage[utf8]{inputenc}
\usepackage[T1]{fontenc}
\usepackage{lineno}
\usepackage{graphicx}
\usepackage{subcaption}
\usepackage{caption}
\usepackage{natbib}
\usepackage{hyperref}
\usepackage{graphicx}
\usepackage{color}
\usepackage{array}
\usepackage{dcolumn}
\usepackage{bm}
\usepackage{multirow}
\usepackage[version=4]{mhchem}
\usepackage{cleveref}
\usepackage{booktabs}
\usepackage{tikz}
\usepackage{sidecap}
\usepackage{pifont}

\newcommand{\ETH}{Department of Materials, ETH Z\"urich, Z\"urich, CH-8093, Switzerland}
\begin{document}

\title{Machine-Learning-enabled ab initio study of quantum phase transitions in SrTiO$_3$}
\author{Jonathan Schmidt}
\affiliation{\ETH}
\author{Nicola A. Spaldin}
\affiliation{\ETH}
\begin{abstract}
   We use the self-consistent harmonic approximation (SSCHA) with machine learning interatomic potentials to calculate the effect of $^{18}$O substitution on the properties of quantum paraelectric SrTiO$_3$ (STO). We find that calculations including both quantum and anharmonic effects are able to reproduce the experimentally observed isotope effect, in which replacement of $^{16}$O by $^{18}$O induces the ferroelectric state, and demonstrate that the ferroelectric phase transition in ST$^{18}$O can be reproduced in a purely displacive manner.  We calculate the ferroelectric soft mode frequency as a function of volume, lattice parameters and temperature for ST$^{16}$O and ST$^{18}$O, and find that the phase space in which ST$^{16}$O shows quantum paraelectric behaviour, while ST$^{18}$O becomes ferroelectric is narrow. Our study shows that machine learning interatomic potentials enable temperature-dependent simulations that include quantum and anharmonic phonon effects, however quantitative prediction of phase diagrams remains challenging due to a lack of universally accurate electronic structure methods.  
   
\end{abstract}
\maketitle
\section{Introduction}

Perovskite-structure strontium titanate, SrTiO$_3$ (STO) is the prototypical quantum paraelectric material, in which suppression of incipient ferroelectricity occurs at low temperature and is attributed to quantum fluctuations of the nuclei \cite{Muller/Burkard:1979}.
The signatures of quantum paraelectricity are well established: A high static dielectric constant \cite{Weaver:1959} with $\frac{1}{T^2}$ scaling \cite{Rowley_et_al:2014} at low temperature, accompanied by a transverse optical phonon that softens but does not reach zero frequency \cite{Cowley:1962}. However, a detailed characterization of the quantum paraelectric state is challenging, particularly in STO, where additional structural complexity is caused by an antiferrodistortive phase transition (AFD) at $\sim$ 100K \cite{Yamada1969} which lowers the symmetry from cubic to tetragonal. In addition, a true ferroelectric state can be induced by small stress- \cite{Uwe/Sakudo:1976,Haeni_et_al:2004} or chemistry-driven \cite{Bethe/Welz:1971} increases in lattice constant, as well as by oxygen isotope substitution \cite{Itoh1999}. 

The oxygen isotope effect, in which ferroelectricity is induced by replacing the usual $^{16}$O isotope fully or partially by $^{18}$O, is particularly intriguing. It was first identified by Itoh and coworkers~\cite{Itoh1999} through polarization hysteresis loops and a peak in the dielectric constant at the Curie temperature, $T_c \sim 20$K for full substitution, with partial substitution down to a critical $^{18}$O concentration $x_c \sim$ 33 \% yielding progressively lower Curie temperatures \cite{Itoh2000}. Subsequent studies using diverse techniques including second harmonic generation \cite{Zhang2002}, Raman spectroscopy \cite{Abe_et_al:2002, RischauSTO}, nuclear magnetic resonance (NMR) \cite{Blinc2005}, and polarized light scattering \cite{Takesada2006} confirm the low-temperature ferroelectric behavior, and a <110> polar orientation is now established \cite{Shigenari_et_al:2006}. The detailed natures of the phase transition and the low-temperature state remain controversial, however. The persistence of polarization -- electric field hysteresis above $T_c$ and the absence of specific heat anomaly \cite{wang2000} point to an order-disorder transition with regions of broken symmetry already above the critical temperature. This behavior is supported by Raman spectroscopy, which indicates symmetry breaking suggestive of polar clusters both above $T_c$ \cite{Yagi2002,Yagi2002b} and below $x_c$ \cite{Taniguchi2006}, as well as NMR, which reveals polar clusters tens of kelvins above $T_c$ that freeze out and percolate at the transition \cite{Blinc2005}. Polarized light scattering, however, points to a reassignment of the soft modes, which, if correct, would imply ideal displacive behavior \cite{Takesada2006}. The critical exponents extracted from temperature dependence of the polarization and dielectric susceptibility range from classic mean-field-like values corresponding to displacive behavior \cite{filipic2006critical, PhysRevB.75.027102} to  fluctuation-dominated \cite{Kleeman/Dec:2007} depending on the temperature range and model used for the fitting.  
To further complicate the situation, in thin films, which are also affected by strain, the transition is reported to be first order  \cite{Fujii2011}. Interestingly, isotope substitution also strongly affects the critical temperature of the superconductivity that occurs in electron-doped samples \cite{Binnig_et_al:1980}, suggesting that ferroelectric fluctuations are relevant to the superconducting pairing mechanism \cite{edge2015,Stucky_et_al:2016,RischauSTO}. 
The resolution of these discrepancies calls for a first-principles study, incorporating anharmonic thermal effects and with a fully quantum-mechanical treatment of both electrons and nuclei, of the para- and ferro-electric behaviors of ST$^{18}$O.

{\it Ab initio} approaches for calculating temperature-dependent anharmonic processes have been established for over a decade~\cite{Souvatzis_et_al:2009}, and have been successfully applied to studying the high-temperature thermal conductivity in SrTiO$_3$ \cite{Han_et_al:2023}. However, their computational cost is prohibitive when incorporating nuclear quantum effects. Effects that are essential for a realistic description of STO, as their omission incorrectly yields a ferroelectric ground state for $^{16}$O.  
Until recently, therefore, the preferred approach was to combine path-integral Monte Carlo (PIMC) simulations, which allow simultaneous treatment of temperature, anharmonicity, and nuclear quantum effects, with an effective Hamiltonian fit to first-principles data. This methodology was used to calculate the effects of quantum fluctuations on the structural phase transitions in SrTiO$_3$ \cite{Zhong/Vanderbilt:1996}, revealing that they completely suppress the ferroelectric transition, consistent with earlier assumptions. In addition, the nature of the fluctuations was found to evolve with temperature, with the high-temperature thermal fluctuations resembling the soft eigenvector of the force-constant matrix, in contrast to the low-temperature quantum fluctuations that emphasize the low-mass oxygen atoms. 

Theoretical studies of the isotope effect to date have relied on phenomenological models with various levels of approximation. 
Typical phenomenological Hamiltonians include generalized quasiharmonic models for double Morse-type local potentials~\cite{tchoubiap}, nonlinear polarizability models~\cite{BussmannHolder2006}, and models for electronic excitations in TiO$_6$ octahedra~\cite{PhysRevB.69.024103}. 
While the double Morse-type local potential used in Ref.~\cite{tchoubiap} captures the temperature-dependence of the soft-mode frequencies, the parameters in the potential have to be adjusted with oxygen isotope concentration to obtain the transition to the ferroelectric state. The nonlinear polarizability model in Ref.~\cite{BussmannHolder2006} was fitted to the experimental phase transition temperature and observed the formation of local polar nano-domains above T$_\text{C}$, which subsequently freeze out below T$_\text{C}$ without establishing long-range order.
The three-state model for electronic excitations in Ref.~\cite{PhysRevB.69.024103} was also able to reproduce macroscopic and microscopic effects of the isotope replacement.

Newer work combines DFT calculations of the soft-mode potential energy surface, with an auxiliary single-particle nuclear Schr\"odinger equation. Using this approach, two groups were able to obtain a quantum paraelectric low-temperature phase~\cite{tobiasSTO, rubiosto}. However, the isotope effect was found to be unphysically small~\cite{tobiasSTO}.  While these models can provide interesting insights, they rely on already available experimental data and/or do not capture all the relevant physics, and our objective is to make fully ab-initio predictions.

In the last few years, two methodological developments have transformed the landscape for studying anharmonicities and quantum effects in solids. First, methods have been developed that calculate vibrational properties of materials with full quantum and anharmonic effects via minimization of the temperature-dependent free energy in terms of parametrization of the N-body density matrix of nuclei. Examples are the stochastic self-consistent harmonic approximation (SSCHA)~\cite{SSCHA} and quantum self-consistent ab-initio lattice dynamics (QSCAILD)~\cite{qscaild} methods, both of which have public implementations and interface with standard density functional theory packages for the ab initio component. 

Second, the use of machine learning interatomic potentials (MLIPs) has revolutionized the field of molecular dynamics, allowing ab initio quality calculations at classical cost assuming a sufficiently well trained MLIP. A number of recent studies have combined SSCHA or QSCAILD and MLIPs to reproduce the quantum paraelectric low-temperature phase in STO and related materials~\cite{STO,kapitza,borissto,KTaO3_1,KTO}. Particularly relevant for us is a study by Verdi et al.~\cite{STO}, who obtained a quantum paraelectric ground state for STO at its experimental volume using MLIPs and the SSCHA method~\cite{STO}. 

In this article, we leverage this new combination of capabilities to study the isotope effect in quantum paraelectric STO. Using a combination of MLIPs and the SSCHA method, we calculate the full phase diagram of STO as a function of volume, degree of tetragonality, oxygen isotope mass, and temperature. Our main finding is that quantum treatment of the atomic nuclei combined with anharmonic description of the lattice vibrations reproduces qualitatively the experimentally observed isotope effect in STO. Our study of this particularly delicate behavior also highlights the capabilities and limitations of MLIPs combined with SSCHA.
\begin{figure}
    \centering
    \includegraphics[width=0.45\linewidth]{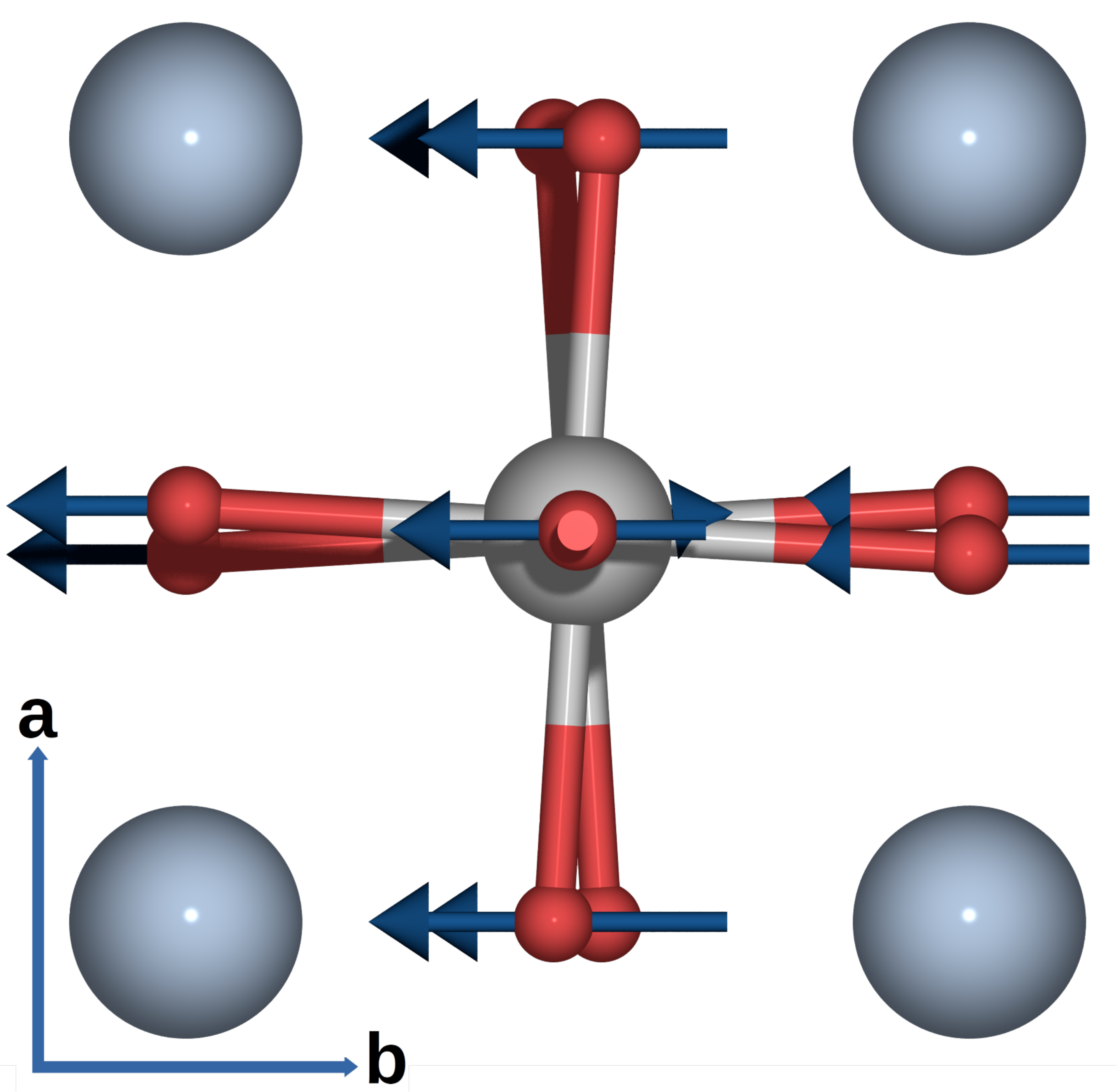}
    \includegraphics[width=0.45\linewidth]{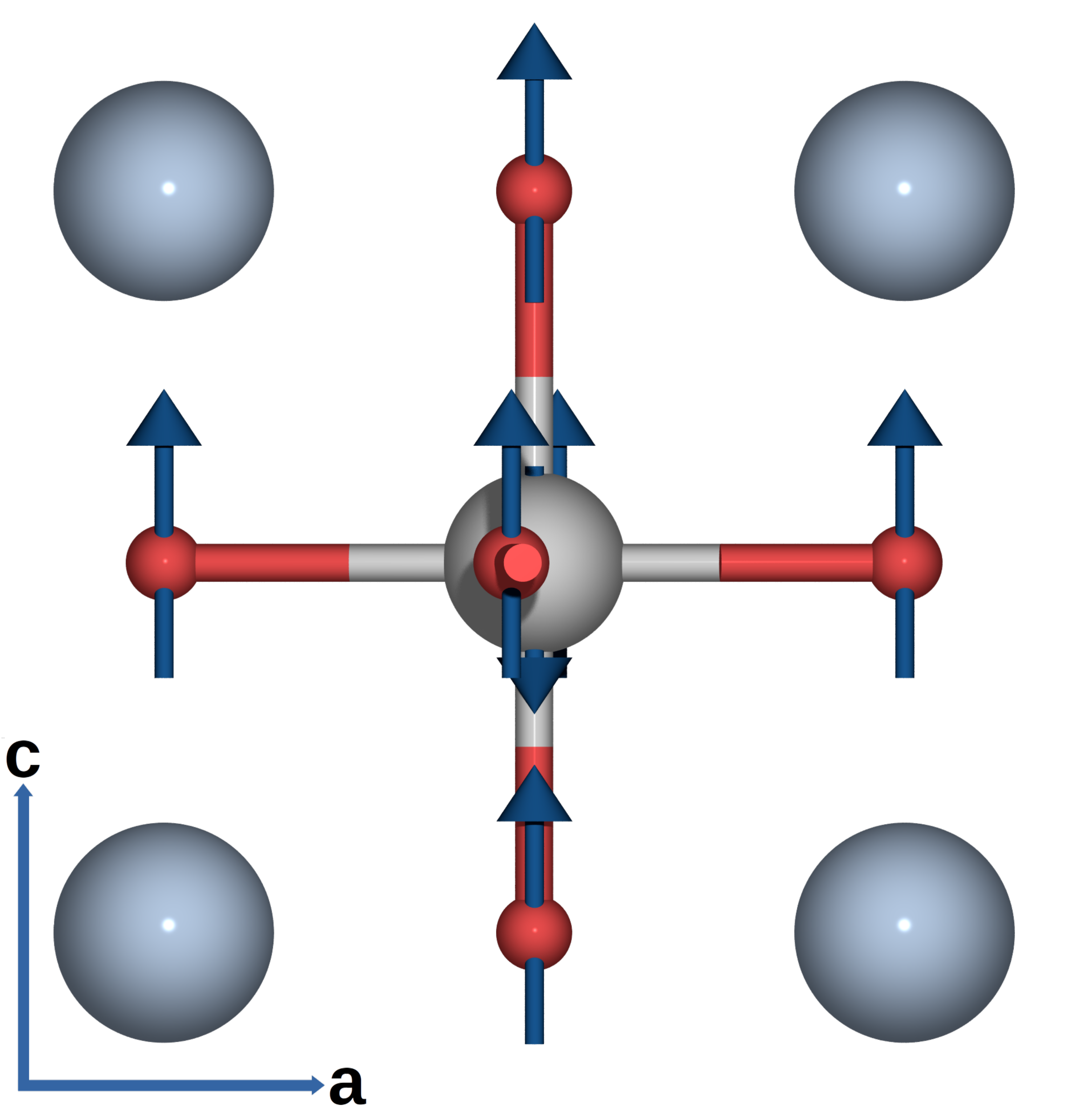}
    \caption{The arrows represent the movement directions and relative amplitude of the atoms for one of the E$_\text{u}$ modes (left) and the A$_\text{2u}$ mode (right). The second E$_\text{u}$ mode is oriented perpendicular.}     
    \label{fig:mode-visualization}
\end{figure}

It is well established that the structural and dynamical properties of STO are highly sensitive to the flavor of DFT, in particular the choice of exchange-correlation functional~\cite{Wahl/Vogtenhuber/Kresse:2008,Aschauer/Spaldin:2014}. Verdi et al.~\cite{STO} obtained a quantum paraelectric ground state for STO at its experimental volume using the random phase approximation (RPA)~\cite{Kaltak2014}. 
They established that RPA  in combination with SSCHA provides an accurate description of the quantum paraelectric soft-mode behavior of ST$^{16}$O while an MLIP trained on strongly constrained and appropriately normed (SCAN) functional provides a better description in terms of structural parameters and the cubic-to-tetragonal phase transition.
As we are interested in the quantum paraelectric and ferroelectric ST$^{18}$O we focus our simulations on RPA-MLIP calculations.
An in-depth discussion and evaluation of the RPA-MLIP can be found in Ref.~\cite{STO}. Verdi et al.~\cite{STO} used a delta-learning approach, training a first MLIP on cheap PBEsol data, a second MLIP was trained  on the differences between the PBEsol MLIP and a limited set of RPA calculations, and finally a last MLIP was trained to reproduce the sum of the two models to accelerate the calculations. To enable compatibility with a new \textsc{vasp} version 6.4.0~\cite{vasp1,vasp2, vaspmlff} we refitted the RPA MLIP using the parameters and data from Ref.~\cite{STO}.

SSCHA approximates the ionic N-body density matrix using a Gaussian distribution, and therefore cannot fully capture the true density matrix for systems where tunneling between, for example, the minima of a polar double-well potential plays a significant role.  However, SSCHA can accurately reproduce the expected atomic positions of such systems~\cite{SSCHA}, which we expect to be sufficient to model the isotope effect in STO.

As mentioned, STO has the cubic perovskite structure at temperatures above 105\,K, below which it transitions to a tetragonal phase with an antiphase tilting of the octahedra, resulting in a doubling of the unit cell.

The tetragonal phase has low-frequency polar vibrational modes of E$_{\text{u}}$ (doubly degenerate in plane) and A$_{\text{2u}}$ (along the $c$ axis) symmetry displayed in Fig.~\ref{fig:mode-visualization} that soften as temperature is lowered~\cite{Yamada1969}. In ST$^{16}$O their frequencies saturate at 7.8\,cm$^{-1}$\cite{Yamanaka2000}\,(7.1\,cm$^{-1}$ \cite{inaba1996hyper}) and 16.5\,cm$^{-1}$~\cite{Yamanaka2000} measured down to 7K (2K). 
In ST$^{18}$O, the frequency of the E$_\text{u}$ mode extrapolates to zero at the ferroelectric T$_C$, below which it hardens and becomes Raman active~\cite{Rischau_et_al:2017}. 
In ab-initio calculations within the tetragonal structure, an imaginary frequency of one of these soft modes, and a corresponding energy double-well potential along the soft-mode coordinate, is a signature of a phase transition to a ferroelectric structure. 

Since it is known that RPA strongly overestimates volumes~\cite{Paier2012, Ruan2021, Verdi_et_al:2023}, we calculate the soft-mode frequencies as a function of unit-cell volume and tetragonality (defined by the ratio of the c- to a- lattice constants) for a range of temperatures and the two isotope masses. We will discuss the full phase diagram in Fig.~\ref{fig:Eu}.
Using the MLIP for energy and force evaluations, we perform \mbox{SSCHA} geometry relaxations under fixed non-polar symmetry. The octahedral tilting angles are allowed to relax as part of the simulation. In general, both the RPA and most DFT functionals tend to overestimate these angles. Starting from the harmonic solution in 3 × 3 × 3 supercells of the tetragonal unit cell (270 atoms) with oxygen masses of 16 or 18 a.u., we proceed iteratively as follows: Initially, we run zero-temperature ensembles of 1000 structures; once these have converged, we restart with ensembles of 10000 and then 40000 structures. Upon final convergence, we compute the Hessian --- and hence the soft-mode frequency --- on the converged ensemble if the Kong–Liu ratio exceeds 95 \%. Otherwise, we repeat the calculation with a new ensemble of 40 000 structures. This process allows us to arrive at sufficiently accurate Hessian values at a reasonable cost. The convergence parameters were chosen as 1e$^{-4}$ times the standard deviation of the energy gradients with respect to the auxiliary SSCHA force constant matrix and the centroid positions. 

\begin{figure}[t]
\centering
 \includegraphics[width=0.98\linewidth]{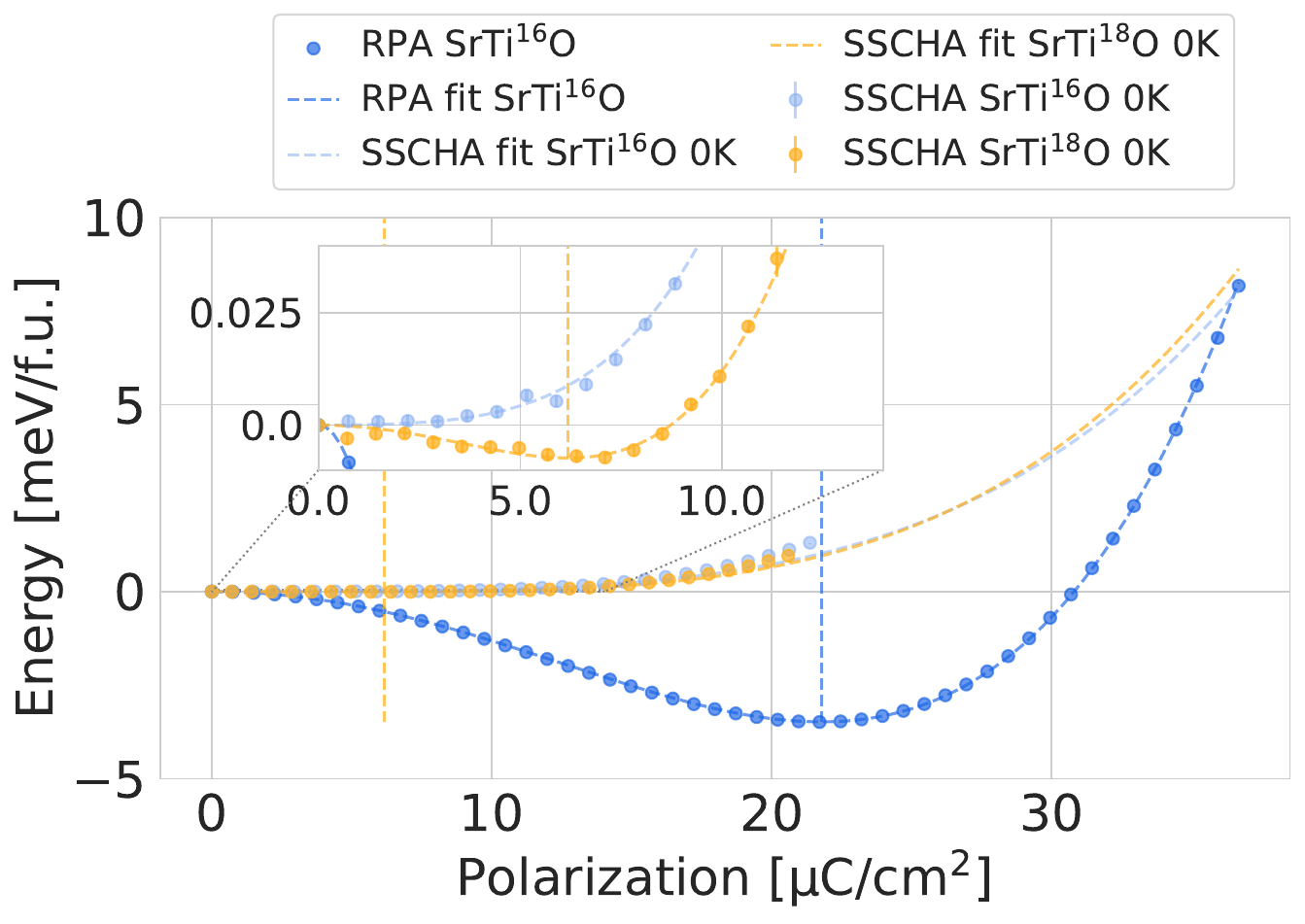}
\caption{Energies per formula unit, relative to that of the non-polar tetragonal structure, calculated using SSCHA for ST$^{18}$O and ST$^{16}$O
      at $c/a=1.0027$ and volume $=59.49\,\text{\AA}^3$, and calculated directly
      with the RPA-MLIP, as a function of polarization.
      All calculations performed at 0\,K. The yellow and blue dashed lines
      indicate the energy minima obtained from the respective double-well fits,
      while the grey dotted line is included as a visual guide for the inset location.}
    \label{fig:double-well}
\end{figure}

\begin{figure*}
    \centering
    \begin{minipage}[t]{0.45\linewidth}
        \vspace{0pt}
        \includegraphics[width=\linewidth]{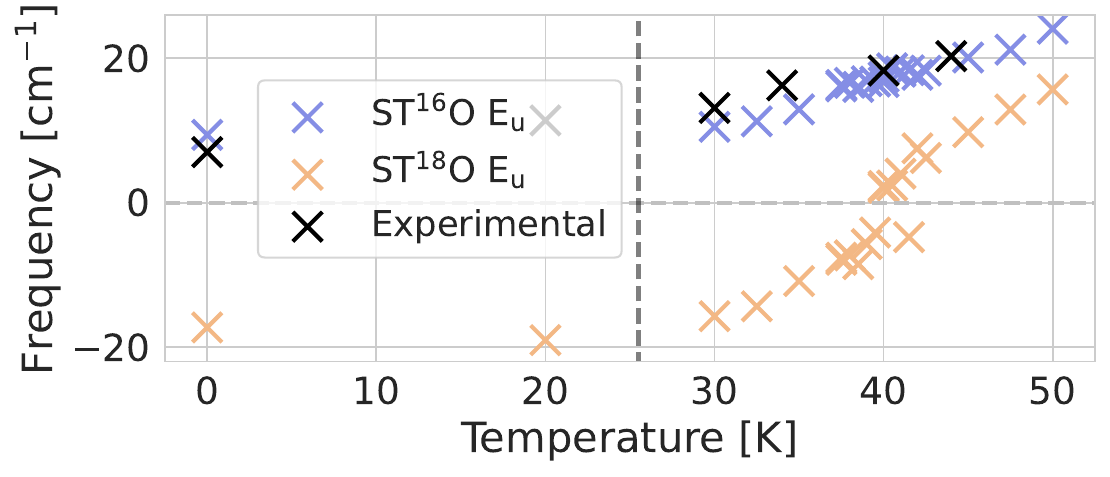}\\[2mm]
        \includegraphics[width=\linewidth]{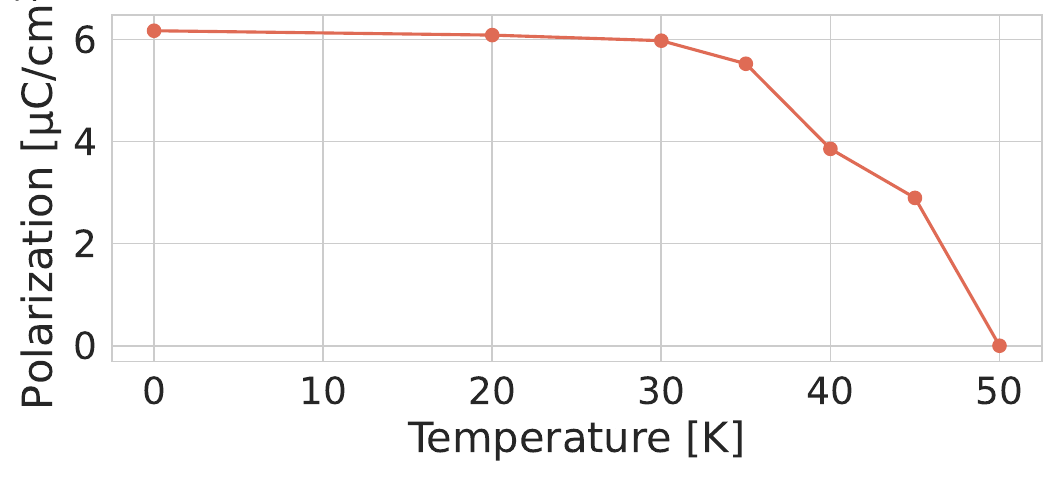}
    \end{minipage}
    \hfill
    \begin{minipage}[t]{0.54\linewidth}
        \vspace{0pt} 
        \includegraphics[width=\linewidth]{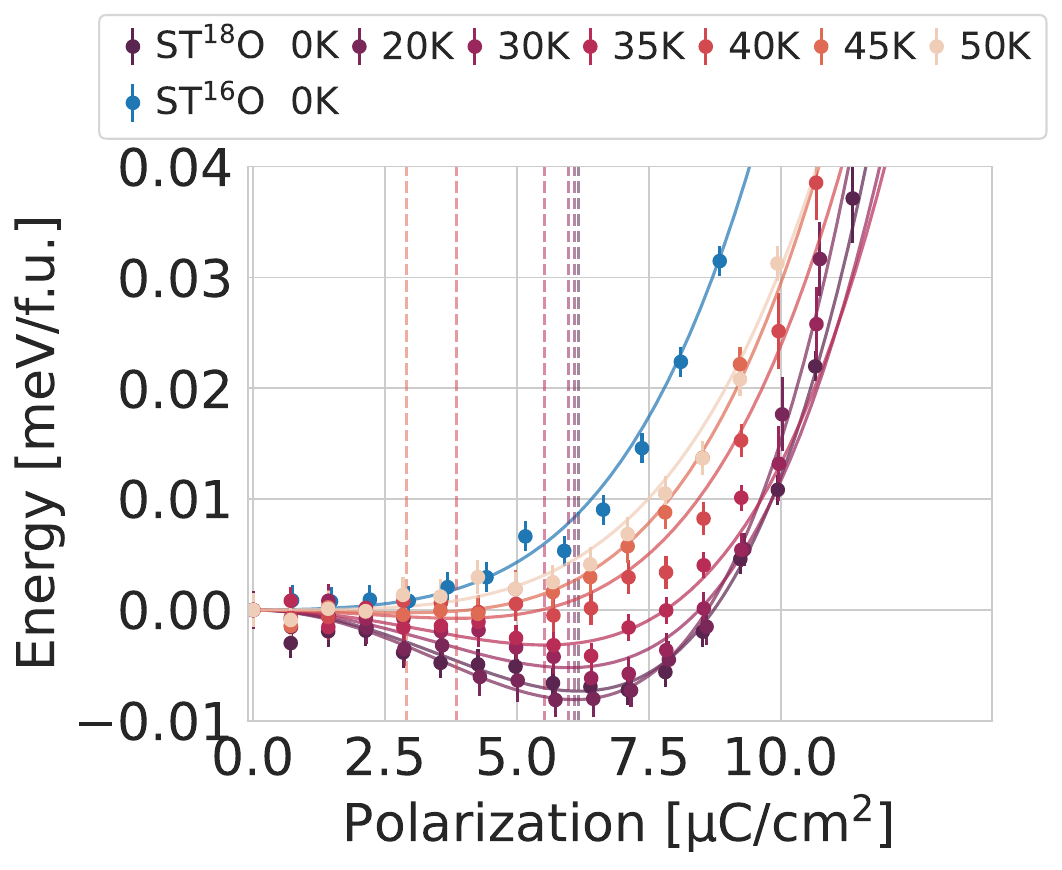}
    \end{minipage}%
    \caption{ In the top left, the E$_\text{u}$ frequency is calculated as a function of temperature for ST$^{16}$O and ST$^{18}$O. Imaginary frequencies are noted as negative and the dashed horizontal line serves as a guide to the eye for the phase transition. The experimental data points with the exception of 0\,K are extracted from Fig.~2 in Ref.~\cite{inaba1996hyper} and the experimental phase transition temperature for ST$^{18}$O is marked by the vertical black line. On the top right, we display the energies per formula unit—relative to the non-polar tetragonal structure calculated using SSCHA for ST$^{18}$O as a function of temperature, along with results for ST$^{16}$O at 0\,K. Vertical lines indicate the fitted minima at each temperature. The bottom left panel displays polarizations as functions of temperature, derived from the data in the right plot.}
    \label{fig:phase-transition}
\end{figure*}

\begin{figure*}[ht!]
    \centering
    \includegraphics[width=0.8\linewidth]{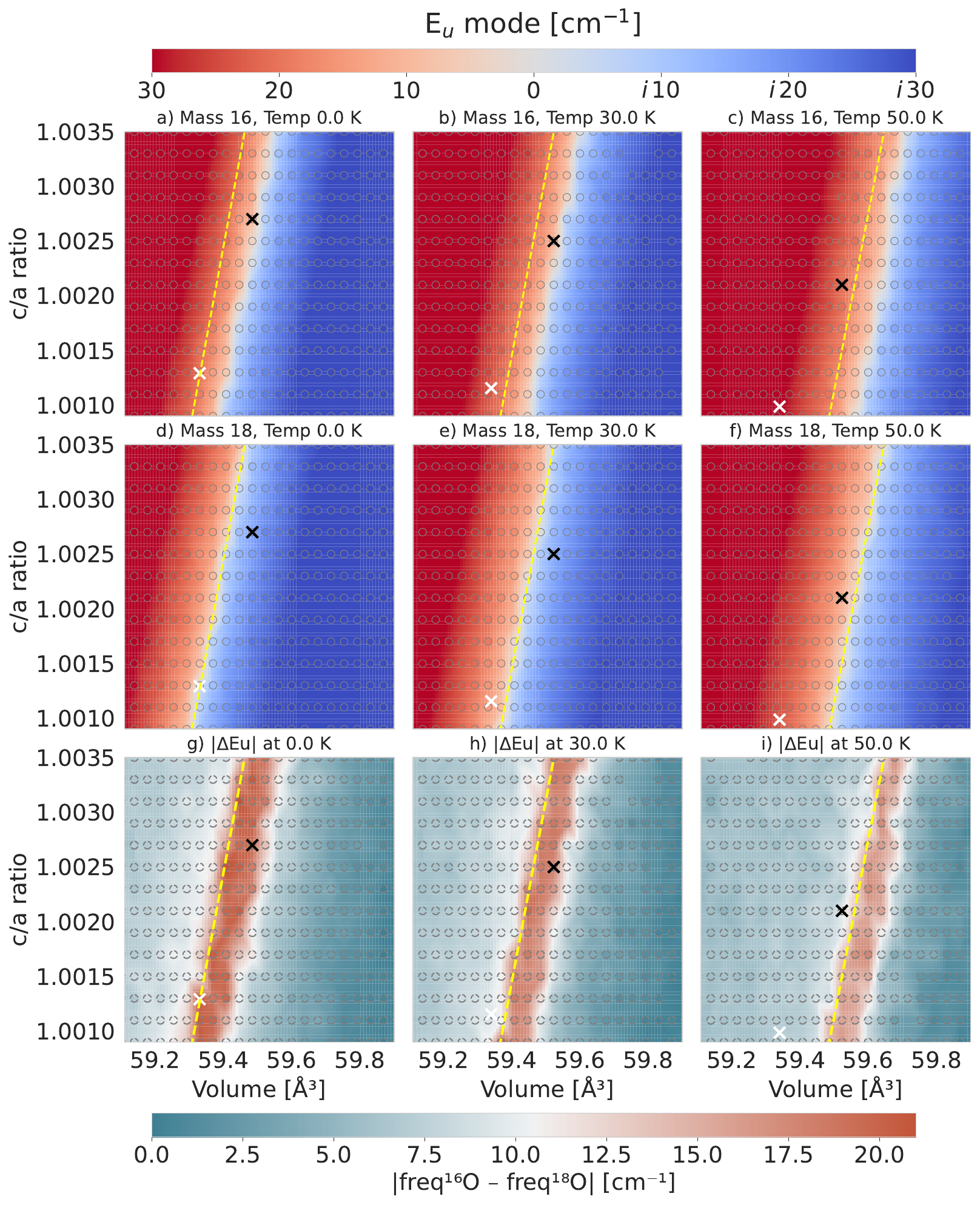}
    \caption{The panels a) to f) display the frequency of the E$_\text{u}$ soft mode calculated with SSCHA at 0\,K, 30\,K and 50\,K. The first row shows the results for $^{16}$O, the second row $^{18}$O and the third row, panels g) to i), illustrates the strength of the isotope effect via the magnitude of the complex frequency difference of the E$\mathrm{u}$ soft mode between $^{16}$O and $^{18}$O. In the first two rows, blue (i.e., imaginary frequencies) indicates ferroelectric structures, while red corresponds to paraelectric structures. The yellow line serves as guide to the eye for the phase transition of ST$^{18}$O at the respective temperature. The white cross marks the experimental structures at the respective temperature (1.5\,K for the zero-temperature diagram~\cite{Kiat1996}) obtained from ICSD~\cite{ICSD}, while the black cross marks the position of the ST$^{16}$O structure with values closest to the experimental E$_\text{u}$ and A$_\text{2u}$ values~\cite{Yamanaka2000}.}
    \label{fig:Eu}
\end{figure*}

We begin by presenting the behavior of the $E_u$ soft mode at 0\,K for the structure, selected from the phase diagram, that yields E$_\text{u}$ and A$_\text{2u}$ mode frequencies closest to the experimental low temperature values for ST$^{16}$O (E$_\text{u}$ 7.1\,cm$^{-1}$\, experimentally vs 10.9\,cm$^{-1}$, A$_\text{2u}$ 16.5\,cm$^{-1}$\, vs 18.9\,cm$^{-1}$)~\cite{Yamanaka2000, inaba1996hyper}. We also confirm for this structure that the SSCHA frequencies saturate at low temperature and only show a slight increase between 0 and 10\,K (see Fig.~\ref{fig:phase-transition}). This structure has c/a=1.0027 and a volume of 59.49\,\AA$^3$, note that these differ from the experimental values measured at 1.5\,K~\cite{Kiat1996}. However, based on the ST$^{16}$O results we expect this structure to provide a similarly good description of the ST$^{18}$O soft mode behavior.

We use the Hessian calculations to freeze in respective SSCHA phonon modes and calculate a double well potential visualized in Fig.~\ref{fig:double-well}. The RPA-MLIP calculations without SSCHA (dark blue line) exhibit a strongly imaginary frequency, leading to a characteristic double-well potential and incorrectly predicting a ferroelectric state --- consistent with previous findings in the literature~\cite{Verdi_et_al:2023}. While the oxygen isotope naturally has no effect on the RPA calculation, it has a small effect on the phonon calculation, as discussed in the supplementary material. Including quantum fluctuations via the SSCHA stabilizes the paraelectric state, rendering the E$_\text{u}$ soft mode real, even though the energy profile as a function of mode amplitude (light blue line) remains strongly anharmonic.
Upon substituting the oxygen atoms with the heavier $^{18}$O isotope, the E$_\text{u}$ mode becomes imaginary, with a frequency of $i$ 17\,cm$^{-1}$. The resulting potential well is considerably shallower (see inset), and the polarization at the energy minimum is much smaller than when quantum effects are neglected. We used an \textsc{atomate2}~\cite{atomate2} DielectricMaker for DFPT in \textsc{VASP} with standard parameters, to compute the Born effective charges of the high-symmetry structure and converted the $E_\text{u}$ mode amplitude to the polarization of the respective modulated structure. This yielded a spontaneous polarization of 6.2\,$\frac{\mathrm{\mu C}}{\text{cm}^2}$ for ST$^{18}$O. For comparison, Filipic et al.~\cite{filipic2006critical} measured the spontaneous polarization in a 94.7\% $^{18}$O‐enriched bulk crystal and  extrapolated to a value of 3.37\,$\frac{\mathrm{\mu C}}{\text{cm}^2}$ at 0\,K.
This first result demonstrates that SSCHA in combination with RPA is able to reproduce the ferroelectric isotope effect in STO and supports the established picture of the origin of the quantum paraelectric behavior and the isotope effect in SrTiO$_3$.

We now identify the structures closest to the experimental E$_\text{u}$ and A$_\text{2u}$ mode frequencies measured at 30\,K and 50\,K in Ref.~\cite{Yamanaka2000} for ST$^{16}$O. We measure the proximity using the sum of squared frequency differences as the distance metric and mark the points as black crosses in Fig.~\ref{fig:Eu}. Using these structures for both ST$^{16}$O and ST$^{18}$O, we calculate a tighter temperature grid in the upper left panel of Fig.~\ref{fig:phase-transition} to determine the phase transition temperature of ST$^{18}$O. The 20\,K value was obtained from an input structure interpolated between the 0\,K and 30\,K structures.
Analyzing the frequencies, we find a phase transition close to 40\,K. This slightly overestimates the experimental phase transition temperature measured at 25\,K~\cite{filipic2006critical} or 26\,K~\cite{RischauSTO}. Using the phonon modes obtained at the various temperatures, we also repeat the SSCHA double well calculation to determine the  polarization as a function of temperature. At 40-45\,K, the double-well fits still predict a finite polarization, but the depth of the double well is already smaller than the uncertainty of the SSCHA energies. By 50\,K, the quadratic term becomes positive, indicating the emergence of a single-well potential.

In Fig.~\ref{fig:Eu}, we present the full phase diagram of the E$_\text{u}$ mode frequency as a function of volume and c/a ratio for three selected temperatures 0\,K (left column) , 30\,K (middle column), and 50\,K (right column) for both isotope masses. The red regions correspond to real phonon frequencies indicating a stable paraelectric state, whereas the blue regions show imaginary frequencies indicating a ferroelectric state. The analogue to Fig.~\ref{fig:Eu} for the out-of-plane A$_\text{2u}$ mode is discussed in the supplementary Fig.~1.

The STO phase diagram shows clear trends with structure, temperature and oxygen isotope mass:
First, at each temperature and for both isotopes, increasing the volume or the a lattice parameter (i.e. decreasing the c/a ratio) favors ferroelectricity, consistent with the experimental response to chemical pressure \cite{Rushchanskii_et_al:2010} or epitaxial strain \cite{Haeni_et_al:2004} as well as earlier DFT calculations~\cite{Esswein/Spaldin:2022, sai:firstprinciple_2000}. The same trends are found using only the RPA-MLIP (see suppl. Fig.\,2). Second, for both isotopes, increasing the temperature disfavors the ferroelectric state, consistent with the usual thermal behavior at a displacive phase transition shifting the transition boundary to larger volumes and c/a ratios. 
Third, at each temperature the ferroelectric phase is stable over a larger region of phase space for the heavier O isotope. 
To highlight the degree of isotope effect, we added the location of the paraelectric / ferroelectric boundary in ST$^{18}$O to the ST$^{16}$O plots as yellow dashed lines, and in the lower panel, figures f) to i), we plot the  magnitude of the complex frequency shifts for the $E_\text{u}$ mode:
\begin{equation}   
  \Delta\omega \;=\;\bigl|\omega_{E_u}^{\mathrm{ST}^{18}\mathrm{O}} \;-\; \omega_{E_u}^{\mathrm{ST}^{16}\mathrm{O}}\bigr|.
\end{equation}
We see that the isotope substitution predominantly influences the frequencies close to the phase transition boundary, consistent with the expectation that perturbations have a more pronounced impact near phase transitions.
 In contrast, plotting the same diagram without SSCHA (see suppl. Fig.\,4) results in a maximum difference of 5 cm$^{-1}$ (3\% max(f$_{\text{Eu-ST}^{16}\text{O}}$, f$_{\text{Eu-ST}^{18}\text{O}}$)), demonstrating the impact of including quantum effects and anharmonicity in the calculations. 

In Fig.~\ref{fig:Eu} the white crosses indicate the volume and c/a ratio of the experimental crystal structures
measured at the respective temperature in Ref.~\cite{Kiat1996}. In panel (a), the zero-kelvin experimental structure of ST$^{16}$O exhibits the expected quantum paraelectric behavior, and panel (d) shows a polar phase lying directly at the phase boundary for ST$^{18}$O.
While the experimental lattice is not fully relaxed in the RPA+SSCHA calculations, the paraelectric/ferroelectric phases for ST$^{16}$O/ST$^{18}$O are nevertheless predicted correctly from a qualitative perspective. Notably, the experimental structures show the same temperature-dependent trend in $c/a$ ratio and volume as the structures selected for their proximity to the experimental frequency values.

With the combined SSCHA+RPA approach we successfully capture the quantum‐paraelectric ground state of SrTi$^{16}$O$_3$ and predict a pronounced isotope‐driven ferroelectric transition in SrTi$^{18}$O$_3$, although it slightly overestimates both the spontaneous polarization and the Curie temperature. 
Machine‐learning interatomic potentials have enabled efficient exploration of the extremely flat soft‐mode energy landscape characteristic of STO, allowing in-principle quantitative predictions of ionic quantum effects  at finite temperature.

However, there are a few technical challenges to our approach that limit its quantitative predictive power.
RPA suffers from a systematic overestimation of the equilibrium lattice volume, which precludes truly quantitative predictions for STO without experimental input. Furthermore, the RPA MLIP is not trained on stresses, complicating SSCHA geometry relaxations that include the cell degrees of freedom. SCAN, on the other hand, predicts equilibrium structures more accurately, but struggles with the description of the quantum paraelectricity~\cite{STO} in STO.

Now, the residual inaccuracies of current electronic‐structure methods for solids remain the principal obstacle to fully predictive modeling: no single functional or many‐body approach simultaneously delivers accurate lattice volumes and soft‐mode behavior in STO. Although existing functionals such as SCAN or more expensive hybrid methods perform well for many materials, challenging systems like STO continue to lie beyond their predictive capabilities.  
A second limitation is the stochastic nature of SSCHA. We  recalculated a large part of the 0\,K and 30\,K phase diagrams for both isotopic masses with different random seeds to estimate the uncertainty in the computed frequencies. We find a median difference of 0.5\,cm$^{-1}$ and a 90th percentile of 1.5\,cm$^{-1}$. However, as shown in suppl. Fig. 3 there are a few outliers in each diagram that can differ by more.
In the supplementary material, we also plot the A$_\text{2u}$ phase diagram and observe that the unstable solutions for the ST$^{18}$O A$_\text{2u}$ mode at the structural coordinates that reproduce the experimental ST$^{16}$O frequencies lie within the uncertainty of the SSCHA calculations at low temperatures.
On the other hand, the selection of structures that reproduce the experimental ST$^{16}$O frequencies is generally stable against the noise, and selecting the next closest points in terms of frequencies would not change our conclusions. We also note that to obtain sufficiently accurate double well potentials, we had to increase the final ensemble size to 100,000 and average over five parallel simulations. Despite that, the error bars corresponding to the standard deviation of the SSCHA calculations are still quite visible in Fig.~\ref{fig:phase-transition} underscoring that the computational acceleration provided by MLIPs is indispensable for reaching the precision required in these calculations. We note that the fitting results for the double wells are sensitive to the maximum polarization included in the fit. Here, we prioritize accurately capturing the double-well region, at the expense of the fit quality at large amplitudes that likely require higher-order terms.

In summary, we have reproduced the experimentally established oxygen isotope effect on the ferroelectric behavior of STO using first-principles calculations. Our study was enabled by MLIPs trained on ab initio DFT calculations and used the SSCHA method to incorporate both thermal and ionic quantum effects self-consistently. Since a recent work introducing quantum ionicity via an auxiliary single-particle nuclear Schr\"odinger equation found an unphysically small isotope effect~\cite{tobiasSTO}, we can conclude that coupled and quantum mechanical electronic / ionic behavior lies at the core of the ferroelectric physics in SrTiO$_3$. Since we do not include disorder in our simulations, our results indicate the feasibility of a standard soft-mode driven phase transition in STO, but do not shed light on the relative roles of displacive and order-disorder contributions. In this context, a recent development coupling machine-learned interatomic potentials with path-integral molecular dynamics to probe quantum disorder in ST$^{16}$O~\cite{Zhu_et_al:2025} suggests a methodological step towards reconciling conflicting experimental results on the nature of the phase transition in ST$^{18}$O. We hope that future studies building on those presented here will also enable improved understanding of the superconducting isotope effect in STO, which is likely connected to fluctuations of the ferroelectric soft mode \cite{edge2015}. Finally, we emphasize that, in spite of the substantial progress enabled by MLIPs combined with SSCHA, accuracy in delicate cases such as STO is limited by the approximations in the underlying electronic-structure calculations, and quantitative predictions should be interpreted with care.  

\section{Acknowledgments}
J. Schmidt, N.A. Spaldin were supported by the European Research Council (ERC) under the European Union’s Horizon 2020 research and innovation program project HERO Grant Agreement No. 810451. Computational resources were provided by ETH Zurich and by the Swiss National Supercomputing Center (CSCS) under project ids s1128 and s1273.

\section{Data Availability Statement}
The MLIP, SSCHA inputs and output structures will be uploaded to Materials Cloud (DOI to be determined). A script to reproduce all plots and fits from the outputs will be uploaded to GitHub, including information to reproduce the computational environment and the versions of libraries such as \textsc{pymatgen}~\cite{pymatgen}, \textsc{ASE}~\cite{ase-paper} and \textsc{phonopy}\cite{phonopy-phono3py-JPCM,phonopy-phono3py-JPSJ} used in pre- and postprocessing.
\bibliographystyle{apsrev4-2}

\end{document}